\documentclass[aps,prl,twocolumn,showpacs,floatfix, superscriptaddress, nofootinbib]{revtex4}
\usepackage{graphicx,bm,hhline}

\newcommand{\intvr}{\int_V\,d^3r\,}

\newcommand{\br}{{\bm r}}

\newcommand{\bcri}{{\bm R}_i}

\newcommand{\nepa}{{n_e^{PA}}}

\newcommand\sss{\scriptscriptstyle}

\begin{document}
\title{Pseudoatom molecular dynamics}
\author{C. E. Starrett}
\email{starrett@lanl.gov}
\author{J. Daligault}
\author{D. Saumon}
\affiliation{Los Alamos National Laboratory, P.O. Box 1663, Los Alamos, NM 87545, U.S.A.}

\renewcommand{\tabcolsep}{4.4pt}

\date{\today}
\begin{abstract}
A new approach to simulating warm and hot dense matter that combines density functional theory based calculations
of the electronic structure to classical molecular dynamics simulations with pair interaction potentials is presented.  The new method,
which we call pseudoatom molecular dynamics (PAMD), can be applied to single or multi-component plasmas.  It gives equation of state 
and self-diffusion coefficients with an accuracy comparable to ab-initio simulations but is computationally much more efficient.
\end{abstract}
\maketitle

The challenge of accurately modeling dense plasmas over a wide range of conditions represents
an unsolved problem lying at the heart of many important phenomena such as inertial confinement
fusion \cite{hammel}, exoplanets and white dwarfs \cite{HEDLP_report, iau147}.  The production of large scale
and accurate tabulations of data such as equation of state and transport coefficients as a function of density and temperature
is a formidable task, requiring a consistent quantum mechanical treatment of the many-electron problem
together with a classical treatment of the nuclear motion.  The atoms in the plasma may have
bound states or be fully ionized, the electrons may be fully degenerate or approaching their classical
limit.  The nuclear fluid can range from weakly through to strongly coupled.  A consistent, reliable
and accurate treatment across all these physical regimes with an approach that remains computationally
tractable remains as an open problem.

Plasmas of interest are typically one to thousands of times solid density, and have temperatures
from about 1eV ($\sim$10kK) to thousands of eV.  The difficulty of creating
and controlling such plasmas in the laboratory explains the 
lack of experimental data to guide theoretical development, though ongoing campaigns at National Ignition
Facility \cite{kritcher14} and elsewhere (eg. \cite{knudson12}), and recent advances in X-ray scattering 
techniques \cite{falk14} are beginning to shed light on this problem.

From a simulations perspective, powerful and complex tools exist that can provide benchmark calculations.  
In the lower temperature regime (a few eV) one such tool is Kohn-Sham (KS) density functional theory molecular dynamics (DFT-MD)
(eg. \cite{desjarlais2}).  Electrons are treated
quantum mechanically through KS-DFT and ions are propagated with classical MD.  The simulations are very computationally 
expensive and this cost scales poorly with temperature, limiting the method to lower temperatures.  In practice KS-DFT-MD
also relies on a pseudopotential approximation, which reduces the computational overhead by limiting the number of actively modeled
electrons, through an ad hoc modification of the electron-nucleus interaction.  
Orbital-free (OF) DFT-MD\footnote{Hereafter referred to as OFMD.} \cite{zerah} does not suffer from the poor temperature scaling of KS-DFT-MD, and it has
been applied to a wide range of plasma conditions (eg. \cite{danel12, clerouin13b}).  This benefit comes at the cost of physical accuracy, though there has been significant
recent progress in improving OFMD towards a KS-DFT-MD level of accuracy (eg. \cite{sjostrom13,sjostrom14}).  However, OFMD remains computationally expensive, with typical
simulations being limited to a few hundred particles and short times.  It too relies on the pseudopotential approximation, so it is not an all-electron
calculation.

\begin{figure}
\begin{center}
\includegraphics[scale=0.5]{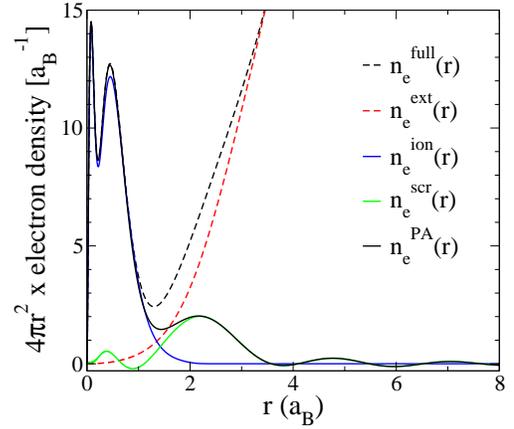}
\end{center}
\caption{(color online)  $4\pi r^2 \times$ electron density for aluminum at 8.1g/cm$^3$ and 1eV.  Shown
are $n_e^{full}(r)$, $n_e^{ext}(r)$ and $n_e^{PA}(r)$, as described in the text.  Also shown is the bound
state (or ion) contribution ($n_e^{ion}(r)$) to $n_e^{PA}(r)$ and the valence electron contribution
$n_e^{scr}(r)$.  The double peak structure in $n_e^{ion}(r)$ reflects the bound state shell structure in 
the aluminum ion, while the oscillations in the valence contribution $n_e^{scr}(r)$ are the well known 
Friedel oscillations, which are damped as temperature increases.  All curves are from the Kohn-Sham version of PAMD.}
\label{fig_nepa}
\end{figure}

Because of this high computational cost, wide ranging equation of state and transport properties tend to
rely on much more approximate methods.  Commonly used techniques include DFT based `average atom' models \cite{feynman, liberman, piron3}, in which one attempts
to solve for the properties of one `atom' in the plasma.  While such models can capture the electronic structure
associated with that atom reasonably well, a consistent treatment of ionic structure resulting in equation of state and 
transport properties of comparable accuracy to DFT-MD has never been successfully included, despite
significant progress towards that goal \cite{ofer, perrot1, rozsnyai14, crowley1, yong09}.  The result is that 
ionic properties, including transport coefficients, are usually calculated more or less independently.  

In this letter we report on a novel method for generating accurate and wide ranging equation of state and
transport properties of dense plasmas, in a single, unified, and internally consistent model.  The method, 
which we call pseudoatom molecular dynamics (PAMD), retains
the computationally efficient average atom approach to the electronic structure of one `pseudoatom',
but couples this with consistent classical MD simulations for the ionic structure, using ab initio
pair interaction potentials.  The vastly reduced computational cost of such calculations relative to DFT-MD
allows for much larger scale simulations.  In short PAMD represents a solution to the problem of consistently 
including ionic structure and dynamics into the average atom methodology.  

Another way to look at PAMD is that it is an approximate version of 
DFT-MD.  The essential approximation is that the plasma can be thought
of as an ensemble of `pseudoatoms' -- this is known as the superposition approximation.  Therefore, PAMD cannot,
for example, accurately model molecules.  However, this limitation is not important for most
of the temperature-density regime discussed above.  The important physics of bound and valence states, 
ion dynamics, as well as ion-ion, ion-electron and electron-electron correlations are all included consistently.
Finally, another important advantage of PAMD over DFT-MD simulations is that it is an all-electron method, 
i.e. no pseudopotential is used.  Not only does this reduce computational complexity, but it removes
uncertainty over possible pseudopotential artifacts.

The key concept of this new method is that of the `pseudoatom' \cite{ziman67, perrot1};  it is
a fictitious, charge neutral object that physically represents a nucleus and its associated electron density, including
bound electrons and its contribution to the valence electrons.  Though its definition is to a certain extent arbitrary,
it was recently shown \cite{starrett13,starrett14} that a satisfactory definition does exist 
and that the pseudoatom electron density $n_e^{PA}(r)$ can be calculated efficiently in a DFT formalism,
using either the orbital-free or Kohn-Sham methods.   In what follows we will show results from both. 
The core idea for calculating $n_e^{PA}(r)$ is to first calculate an electron density $n_e^{full}(r)$ in a system with a nucleus at the origin,
surrounded by a spherically averaged ionic configuration described by the ion-ion pair distribution function $g_{\sss II}(r)$.  
One then calculates the electron density $n_e^{ext}(r)$ in the same system but with the central nucleus removed.  
$n_e^{PA}(r)$ is defined as the difference $n_e^{full}(r) - n_e^{ext}(r)$ (see 
fig. \ref{fig_nepa}).  The physical motivation behind this is to isolate the influence of one nucleus on the electron density.
Furthermore, in \cite{starrett14} is was demonstrated that $n_e^{PA}(r)$ is insensitive to $g_{\sss II}(r)$.
Given this conclusion, one can immediately see that it should be
possible to accurately reconstruct the total electron density $n_e(\br)$ of the plasma as a superposition of pseudoatom
electron densities, each centered at a nuclear site
\begin{equation}
n_e(\br) = \sum\limits_{i}
\nepa(\left| \bcri - \br \right|)
\label{super}
\end{equation}
where $\bcri$ is the position vector of nucleus $i$, and the sum runs over all nuclear sites.
\begin{figure}
\begin{center}
\includegraphics[scale=0.4]{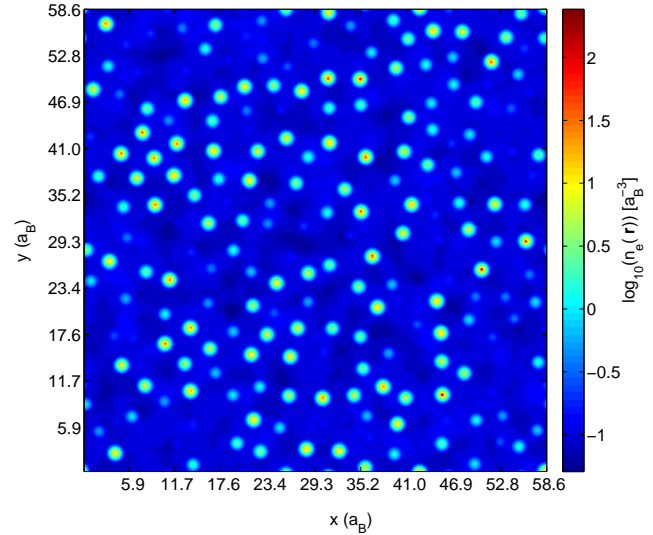}
\end{center}
\caption{(color online) 2-D slice of electron density in the Kohn-Sham version of the model for aluminum 
at 8.1g/cm$^3$ and 1eV.  The plot shows log$_{10}$ of the electron electron density.  
For reference log$_{10}$ of the average total electron density is -0.46 and log$_{10}$ of average screening (valence) electron density is
-0.99.  The ion positions were generated in a molecular dynamics simulation with 5000 nuclei using periodic boundary
conditions.}
\label{fig_projection}
\end{figure}

To generate the nuclear configurations $\{{\bm R}_i\}$ we use classical MD with pair interaction potentials in a cubic simulation cell with periodic
boundary conditions, carried out in the micro-canonical ensemble.  An effective 
pair interaction potential between pseudoatoms $V_{\sss II}(r)$ was derived in refs. \cite{starrett13,starrett14}.  In Fourier-space it is
given by\footnote{Here we write the expression for plasmas with one nuclear species, the expression for mixtures is given in reference \cite{starrett14b}.}
\begin{equation}
V_{\sss II}(k) = \frac{4 \pi \bar{Z}^2}{k^2} + \frac{n_e^{scr}(k)^2}{\chi_e(k)}
\label{pair}
\end{equation}
where $\bar{Z} = \int\,d\br\, n_e^{scr}(r)$ and $\chi_e$ is the electron response function \cite{starrett14}.  The screening
density $n_e^{scr}(r)$ is the contribution to the valence electrons from the pseudoatom.  It is defined by first defining
the bound (or ion) states, and calculating their electron density $n_e^{ion}(r)$, so that
\begin{equation}
n_e^{scr}(r) = n_e^{PA}(r) - n_e^{ion}(r).
\label{nescr}
\end{equation}
\begin{figure}
\begin{center}
\includegraphics[scale=0.5]{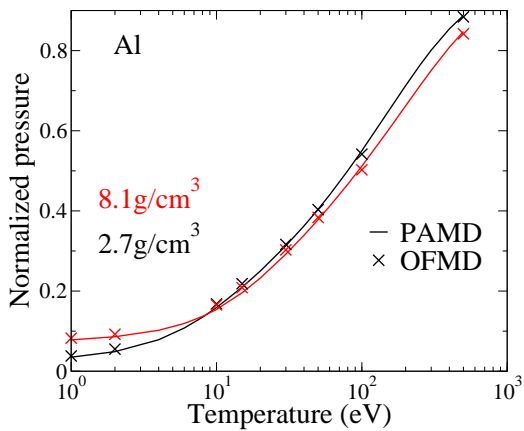}
\end{center}
\caption{(color online) Comparison of pressure for an aluminum plasma between PAMD and OFMD in the Thomas-Fermi approximation.  
We show total pressure divided by the pressure of a fully ionized aluminum plasma of non-interacting classical ions and quantum electrons.
Excellent agreement is found for both densities across this wide temperature range.} 
\label{fig_pal}
\end{figure}

PAMD has no adjustable parameters: the inputs are the nuclear charges, atomic masses, the plasma temperature and mass
density, and a choice of exchange and correlation functional\footnote{For all PAMD and OFMD calculations carried out for this paper
we have used the Dirac exchange functional \cite{dirac} (see also \cite{starrett14}).}.  In fig. \ref{fig_projection}
we show a 2-D slice of the electron density for a Kohn-Sham PAMD simulation with 5000 nuclei, for aluminum at 1eV and 8.1g/cm$^3$.
Each circular object is a slice through a pseudoatom intersecting that plane.  For those pseudoatoms whose nuclei lie closer to the plane in fig. \ref{fig_projection} the
strong localized deformation of the electron density due to the bound electrons is visible.
A simulation of this size would be very challenging for KS-DFT-MD due to computational cost, and will remain so for the foreseeable
future.
\begin{figure}
\begin{center}
\includegraphics[scale=0.275]{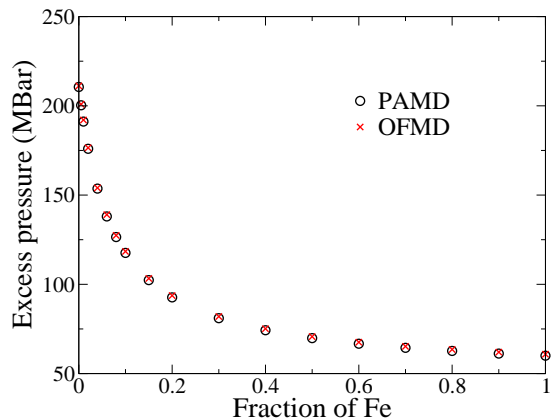}
\end{center}
\caption{(color online) Excess pressure for a mixture of iron and helium at 10g/cm$^3$ and 50eV from PAMD and OFMD \cite{danel09} in the Thomas-Fermi
approximation.  Excess pressure is defined as the total pressure minus the ideal ion contribution (see \cite{danel09}).
We find excellent agreement for all mixing ratios.}
\label{fig_pfehe}
\end{figure}

In finite temperature DFT \cite{mermin65} the grand potential is in principle determined exactly for a given external potential once the electron
density that minimizes it has been found.  Thus, assuming that equation (\ref{super}) is an accurate approximation to the equilibrium electron density for
a given ionic configuration $\{{\bm R}_i\}$, one can determine the thermodynamic properties. For example, in the Thomas-Fermi
approximation the pressure $P$ for a plasma of volume $V$ with $N$ ions and at temperature $k_{\sss B} T$ ($=1/\beta$), 
can be calculated using the virial formula (eg. \cite{yin83})
\begin{equation}
P\,V = N\,k_{\sss B} T + \frac{2}{3} K_e^{TF}[n_e(\br)] +\frac{1}{3} F^{el}[n_e(\br)] + C^{xc}[n_e(\br)]
\label{virial}
\end{equation}
where $K_e^{TF}$ is the Thomas-Fermi approximation to the electron kinetic energy, $F^{el}$ is
the electrostatic free energy and $C^{xc}$ is the contribution from exchange and correlations.  
$K_e^{TF}$ is given by
\begin{equation}
K_e^{TF} = \frac{1}{\beta} \intvr c_{\sss TF} I_{3/2}\left[ \eta(\br))\right]
\label{tfke}
\end{equation}
where $I_{j}$ is the Fermi integral of order $j$ \cite{starrett13} and $c_{\sss TF} \equiv \sqrt{2}\pi^{-2} \beta^{-3/2}$.
The electron density in this approximation is
\begin{equation}
n_e(\br) = c_{\sss TF} I_{1/2}\left[ \eta(\br))\right]
\label{netf}
\end{equation}
Thus $K_e^{TF}$ can be calculated by inverting equation (\ref{netf}) for $\eta(\br)$ and evaluating equation (\ref{tfke}).
$C^{xc}$ and $F^{el}$ are also straightforward to calculate given $n_e(\br)$ from equation (\ref{super}).
In figs. \ref{fig_pal} and \ref{fig_pfehe} pressures calculated from PAMD using equation (\ref{virial}) are compared to 
OFMD simulations in the Thomas-Fermi approximation.  In figure \ref{fig_pal}, for a pure aluminum plasma, agreement is 
excellent throughout the range of temperatures and for both densities.  In fig. \ref{fig_pfehe} we compare pressures for an iron-helium
mixture as a function of the fraction of iron in the plasma.  Agreement is excellent for all iron fractions.

The advantage of using PAMD here is twofold: firstly, no pseudopotential is needed; PAMD is an all electron method. 
Secondly, the calculation proceeds much more quickly.  The calculation of the pseudoatom electron density
and pair interaction potential takes a few minutes on a single processor.  The cost of the classical MD simulations and calculation
of the equation of state depends on the number of particles and the number of time steps.  For the the results presented in fig. \ref{fig_pal}
we used 5000 particles and 40000 time steps; the simulations took $\sim$2.5 hours per point on a single compute node with 24 cores.  Similarly
sized OFMD simulations would be extremely expensive.
\begin{table} \centering
\begin{tabular}{c | c | c | c | c | c }
\hline \hline  Element & $\rho$     & T    & OFMD                     & OFMD & PAMD \\
                       & (g/cm$^3$) & (eV) & \cite{danel12,lambert06} & (This work) & \\
\hline D & 1.5 & 2.5 &  0.0159 & 0.0146 & 0.0154 \\
B     & 1 & 5 &  0.0162 & 0.0156 & 0.0155 \\
B     & 10 & 5 &  0.00240 & 0.00214 & 0.00232 \\
Fe & 22.5 & 10 &  0.0011 & 0.00093 & 0.00105 \\
Cu & 67.4 & 100 & 0.00407 & 0.0039 & 0.00385 \\
\hline
\end{tabular}
\caption{Self-diffusion coefficients $D$ in cm$^2$/s for various element and a range of temperatures ($T$) and densities ($\rho$).  The PAMD
result agrees very well with the OFMD calculations, providing a very sensitive test of the PAMD pair interaction potential.}
\label{tab_diff}
\end{table}

Equation (\ref{virial}) is also valid for Kohn-Sham calculations if $K_e^{TF}$ is replaced by the corresponding 
KS quantity $K_e^{KS}$.  However, one cannot evaluate $K_e^{KS}$ with knowledge of $n_e(\br)$ alone as in the orbital
free case.  Instead $K_e^{KS}$ depends on the Kohn-Sham wavefunctions (orbitals) which are not provided by PAMD.  Approximate 
methods to determine $K_e^{KS}$ in PAMD could be developed but we do not attempt that here.
\begin{figure}
\begin{center}
\includegraphics[scale=0.5]{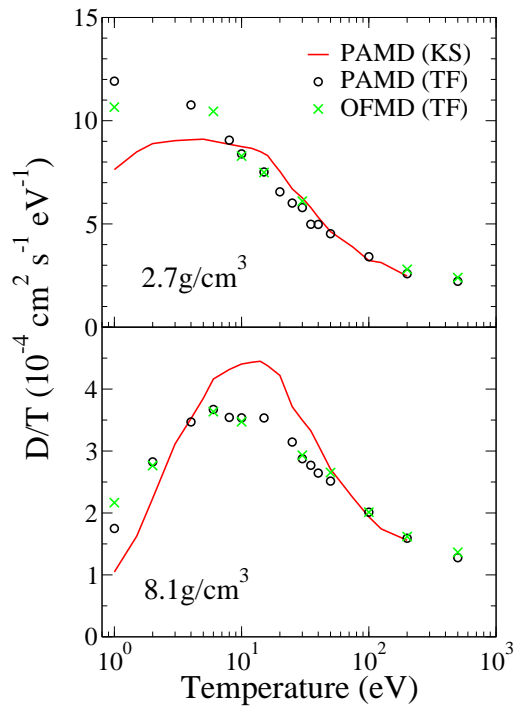}
\end{center}
\caption{(color online) Comparison of the self-diffusion coefficient $D$ for aluminum between PAMD and OFMD in the Thomas-Fermi
approximation.  Also shown is the PAMD calculation using the Kohn-Sham functional from 1 to 200eV.  Such a calculation
would be a formidable task for the ab initio KS-DFT-MD method.  Note that we plot $D$ divided by temperature in eV.}
\label{fig_dal}
\end{figure}

Dynamical ion quantities such as the self-diffusion coefficient $D$, can be calculated with Kohn-Sham or orbital-free
PAMD, since the MD simulations require only the pair interaction potential.  $D$ is calculated using the Kubo relation \cite{hansen1}
\begin{equation}
D = \frac{1}{3}\int\limits_0^\infty \left< {\bm v}(t) \cdot {\bm v}(0) \right> \, dt
\label{kubo}
\end{equation}
where ${\bm v}(t)$ is the velocity of a given ion in the MD simulation at time $t$.
In table \ref{tab_diff} we compare self-diffusion coefficients for a range of materials,
for various densities and temperatures, to published OFMD results \cite{lambert06, danel12} which use
the Thomas-Fermi functional and a range of exchange and correlation functionals.  We have also
repeated these OFMD calculations using the Dirac exchange functional, and these results are also shown 
in table \ref{tab_diff}.  The PAMD results agree very well with the OFMD calculations, for all
the cases.  As a further test, in fig. \ref{fig_dal} we compare the self-diffusion coefficients for 
aluminum from PAMD and OFMD in the Thomas-Fermi approximation.  Agreement is very good for both 
densities and all temperatures.  These comparisons on self-diffusion coefficients represent a very
sensitive test of the quality of the pair interaction potential.  
Such a level of
agreement with an ion dynamical property is quite remarkable, given the very different approaches to the calculation
of ionic forces in PAMD and OFMD.  We also show for comparison in fig. \ref{fig_dal}, the self-diffusion coefficient
as calculated in PAMD using the Kohn-Sham functional.  At the highest temperatures ($>$100eV) there is excellent agreement between the KS and TF
diffusion coefficients.  We see significant deviations from the TF result below $\sim 50$eV
for the higher density but at the lower density agreement between the KS and TF results is reasonable above $\sim 10$eV.  
It is expected that the Thomas-Fermi approximation will be inaccurate for the lower temperatures
due to its ignorance of important quantum effects, that are captured in the Kohn-Sham calculations.  The 
ability of Kohn-Sham based PAMD to quickly evaluate self-diffusion coefficients
across temperature regimes is a significant capability, given the extreme computational cost that 
corresponding KS-DFT-MD simulations would entail.

In conclusion we have introduced a new method to simulate warm and hot dense matter that we call pseudoatom molecular dynamics.  The method
has proved accurate for equation of state and self-diffusion coefficients compared to orbital free molecular dynamics
in the Thomas-Fermi approximation, validating the underlying physical assumption that the plasma can be considered
to be an ensemble of identical pseudoatoms.  The Kohn-Sham version of the model can be applied at high temperatures and calculations of self-diffusion coefficients
for aluminum up to 200eV have been presented.
The low relative cost of PAMD permits wider ranging and larger scale investigations
of the properties of warm and hot dense matter than have hitherto been possible.

This work was performed under the auspices of the United States Department of Energy under contract 
DE-AC52-06NA25396 and LDRD grant number 20130244ER.

\bibliographystyle{unsrt}
\bibliography{phys_bib}

\end{document}